# Detection of Nuclear Sources in Search Applications using Dynamic Quantum Clustering of Spectral Data

Marvin Weinstein[1,2], Alexander Heifetz[3*], Raymond Klann[3],
[1]Quantum Insights LLC, Palo Alto, CA, 94303
[2]Stanford Linear Accelerator Center (Emeritus), Stanford, CA, 94025
[3]Nuclear Engineering Division, Argonne National Laboratory, Lemont IL 60439
*Corresponding author

**Abstract**
In a search scenario, nuclear background spectra are continuously measured in short acquisition intervals with a mobile detector-spectrometer. Detecting sources from measured data is difficult because of low signal to noise ratio (S/N) of spectra, large and highly varying background due to naturally occurring radioactive material (NORM), and line broadening due to limited spectral resolution of nuclear detector. We have invented a method for detection of sources using clustering of spectral data. Our method takes advantage of the physical fact that a source not only produces counts in the region of its spectral emission, but also has the effect on the entire detector spectrum via Compton continuum. This allows characterizing the low S/N spectrum without distinct isotopic lines using multiple data features. We have shown that noisy spectra with low S/N can be grouped by overall spectral shape similarity using a data clustering technique called Dynamic Quantum Clustering (DQC). The spectra in the same cluster can then be averaged to enhance S/N of the isotopic spectral line. This would allow for increased accuracy of isotopic identification and lower false alarm rate. Our method was validated in a proof-of-principle study using a data set of spectra measured in one-second intervals with Sodium Iodide detector. The data set consisted of over 7000 spectra obtained in urban background measurements, and approximately 70 measurements of Cs-137 and Co-60 sources. Using DQC analysis, we have observed that all spectra containing Cs-137 and Co-60 signal cluster away from the background.

1. Introduction

In a search scenario, nuclear background spectra are continuously measured in short acquisition intervals (usually one second) with a mobile detector-spectrometer (i.e. in a vehicle or carried in a backpack). In principle, sources can be detected and identified from the measured data by their unique spectral lines. Detecting sources from data measured in a search is difficult because of low signal to noise ratio (S/N) of spectra, large and highly varying background due to naturally occurring radioactive material (NORM), and line broadening due to limited spectral resolution of radiation detectors.

Data analysis is typically performed in post-processing mode, and consists of saliency and anomaly searches in a data set containing tens of thousands of one-second spectra. Visual inspection of data by a human analyst is not efficient because of large volume of data. Existing computational tools utilize a total counts threshold-triggered alarm, and radiation isotope identification (RIID) algorithms. Using threshold-based alarm results in a large number of false alarms due to large background fluctuations in an urban search, and may result in missed detection of a weak source with sub-threshold total counts. RIID algorithms applied to one-second spectra produce a large number of false alarms because of statistical background fluctuations. The number of false alarms can be decreased if one second-spectra are averaged over some time window to improve the S/N. However, the size of the window is not known a-priori. Averaging over the wrong window would decrease the S/N of a source by washing out spectral lines with random counts. In general, the spectra containing measurements of the same source might not be sequential in time. This can happen if the detector approaches the same sources at different times during the search, or if the source has been moved. Since there is no a-priori information, one can average over randomly selected combinations of one-second spectra. Given the large size of the data, the solution of such a combinatorial problem would require enormous computational resources.

We have developed a method for the detection of sources in search applications using a data clustering technique called Dynamic Quantum Clustering (DQC) [1]. Our approach provides the scientific basis for grouping one-second spectra of the same source, irrespective of time stamps of these spectra. The physical idea behind this approach is that the presence of a source would not only impact a narrow spectral band corresponding to emission line, but also has the effect on the entire detector spectrum via the Compton continuum. This allows characterizing spectra, including the ones without distinct isotopic lines, using multiple data features. DQC selects the spectra according to their overall shape similarity. The cluster or clusters possibly containing source can be identified in the feature space by their separation from the rest of the data. The spectra in the same cluster can then be averaged to enhance S/N of the isotopic spectral line. This would allow for increased accuracy of isotopic identification and a lower false alarm rate.

2. **Challenges of Nuclear Detection in Search Applications**

Detection of radioactive isotopes during a nuclear screening campaign consists of performing both saliency and anomaly searches in large amounts of measured spectral data. In a typical drive-through search scenario gamma spectra are continuously measured in air over short acquisition time intervals, (e.g., one second), with a fast response nuclear detector-spectrometer, such as Thallium-doped Sodium Iodide (NaI(Tl)) scintillator[1-4]. The NaI(Tl) detector is typically set up with 1024 or 512 spectral channels to measure an energy range from 0 to 3MeV. Each radioactive isotope has unique emission lines. In principle, the presence of a particular isotope can be inferred from the measured gamma spectrum either by an experienced spectroscopist, or by using a digital computer running automated radioisotope identification (RIID) algorithm. The challenges of nuclear spectroscopy in a search scenario, which set it apart from the conventional spectroscopy, consist of (1) low signal-to-noise (S/R) ratio of gamma-ray spectral lines; (2) large and highly varying background; and (3) line broadening due to limited spectral resolution of nuclear detector.

Strength of the signal in nuclear detection depends on a number of factors, including the activity of the source, source to detector distance, presence of source shielding materials, and the detector size and efficiency. For example, gamma photon flux from an isotropic point source decreases as the square of the distance. In addition, photon flux undergoes absorption and scattering losses in air. To quote the order of magnitude, for photons of E=600keV, the mean free path in air is approximately 100 m. Source shielding materials, if present, result in further attenuation of the flux. In case of low counts, S/N can be improved by increasing the integration time. However, in a wide area search scenario, the spectral data is typically acquired in a sequence of one-second measurement intervals due to moving detectors and the need to screen large areas quickly. Because of the probabilistic nature of the source to detector distance and the short acquisition time during screening, most measured spectra have low signal-to-noise (S/R) ratios. The signals of interest are weak even for sources without shielding, and might not exhibit distinctive isotopic spectral lines. Within the framework of a particular radioactive background, nuclear detection is associated with measurement uncertainties. For example, in each energy bin in the spectrum, gamma count as a function of time has a Poisson distribution, or degrades to a Binomial distribution for very small sample rates. Because there is limited measurement time available in a search scenario, there will be a large temporal variance in the amplitude of the detected signal spectrum. Thus the potential increase in the count rate for a particular energy bin - due to the presence of special nuclear material (SNM) or other radioactive materials- can be smaller than the variance of the background count in that bin.

Another complication for nuclear search stems from the presence of a fairly strong and highly varying nuclear background due to emission by naturally occurring radioactive materials (NORM). The radioactive background changes randomly due to a number of natural and man-made causes. These include but are not limited to the presence of building materials (functioning both as source and shielding), variation in concentration of NORM isotopes in the soil, increasing flux of the cosmic background radiation with increasing elevation above the sea level, and seasonal and nocturnal variations. The nuclear background spectrum in the energy bandwidth from 0 to 3MeV and measured with a NaI(Tl) detector is dominated by gamma emission of primordial radioactive isotopes K-40, U-238, Th-232,, as well as daughter products of natural decay series of the U-238 and Th-232. While these primordial isotopes are found in virtually all surroundings on Earth, urban building materials typically contain high concentrations of NORM. This frequently causes large random fluctuations in background count in an urban search. In addition to random background variations and noise, isotopic detection is further complicated by potential shielding. If a shielded SNM source is present at the scene, unknown shielding would result in unpredictable attenuation and distortion of the isotopic spectra. Thus, the anomaly to search for in the data is not well defined ahead of time.

Additional challenges in nuclear spectrum analysis come from line broadening due to the limited spectral resolution of nuclear detectors. NaI(Tl) is one of the most commonly used gamma ray detectors because of a high response speed, relatively low cost, and technological ability to produce large volume NaI(Tl) crystals [4]. However, a NaI(Tl) detector has spectral resolution of approximately 6-7% at 662 KeV (~46KeV FWHM). Thus, spectra lines that are theoretically expected to be narrow appear as broad Gaussian shaped peaks when measured with NaI(Tl) detector. This is because of the statistical fluctuations in the number of free charge carriers excited per monoenergetic interaction event in the scintillator crystal. In addition, non-zero counts, with particularly high number contribution in the low energy range, are registered below spectral peaks. These counts are referred to as Compton continuum. They appear due to scattering events taking place in the scintillator crystal where only partial energy deposition occurs because the scattered gamma-ray escapes the detector crystal.

3. **Current Approaches to Nuclear Search Data Analysis**

A screening campaign usually lasts for several days, with 8-12 hour shifts. During each day, there are approximately 30,000 one-second spectra per detector platform to analyze. Current practices employed by radiological response teams consist of monitoring the count rate levels for large anomalies in real-time (i.e. threshold alarms) and analyzing nuclear background data collected during the work-day in off-line mode at the end of the shift. The usual approach to data analysis consists of examining time-series of spectrum-integrated total counts. If during visual inspection of time-series the analyst discovers what he or she believes to be an anomaly, spectral content of the suspicious data is subjected to further examination. Most current nuclear detection systems are coupled with GPS receivers. If an anomaly is found in the spectral data, then one would return to the corresponding geographical location to make additional measurements. Although real-time detection of sources is the ultimate goal, with the current state of technology, data analysis in a post-processing mode results in more thorough review of the data and higher accuracy of detection.

### 3.1 Computational Methods using Total Counts Threshold Alarm

Because of a high probability of human error in analyzing large volume of nuclear background data, utilizing computational tools has the potential of increasing the accuracy of detection. In addition, using a computer-based method brings the possibility of extending background data analysis to real-time mode. Most commercial systems use an algorithm which produces an alarm if the total counts, or the counts within a spectral region of interest, exceed some threshold value. If an alarm is triggered, spectral content of the data is examined.

The problem is that using an alarm trigger - based on some threshold value of total counts - results in large number of false alarms and missed detections. This is because abrupt changes in the measured background rate, such as when a mobile platform detector moves between buildings, might accidentally trigger a false alarm. On the other hand, a source which does not produce enough counts above the threshold would not trigger an alarm.

To illustrate the problems that occur, let us consider a specific example. Figure 1 shows the trajectory of a drive through a portion of downtown Chicago with an Exploranium GR-460 commercial NaI(Tl) crystal-based 512-channel detector. Background spectra were measured every second. The plot shows that the gamma exposure rate due to background varies significantly depending on location. (It also shows intermittent GPS signal due to urban canyons.) A histogram plot of the gamma count rate is shown in Figure 2. The count rate varies widely from a low of about 1000 cps in the park areas with minimal structures to as high as 5000 cps near many of the buildings. These significant changes are all due to background NORM materials.

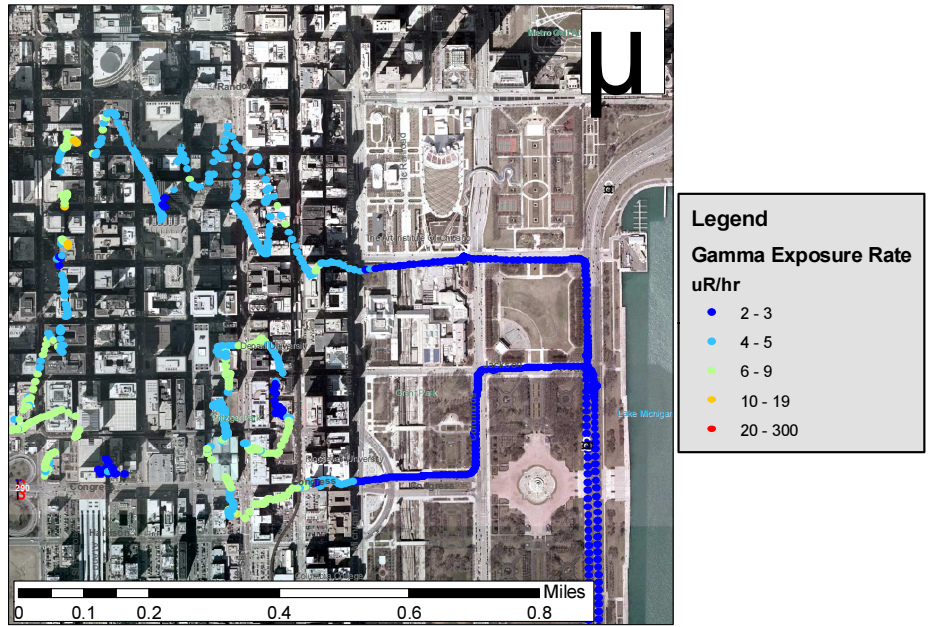

Figure 1: Gamma Exposure Rate (µR/hr) in a section of Chicago.

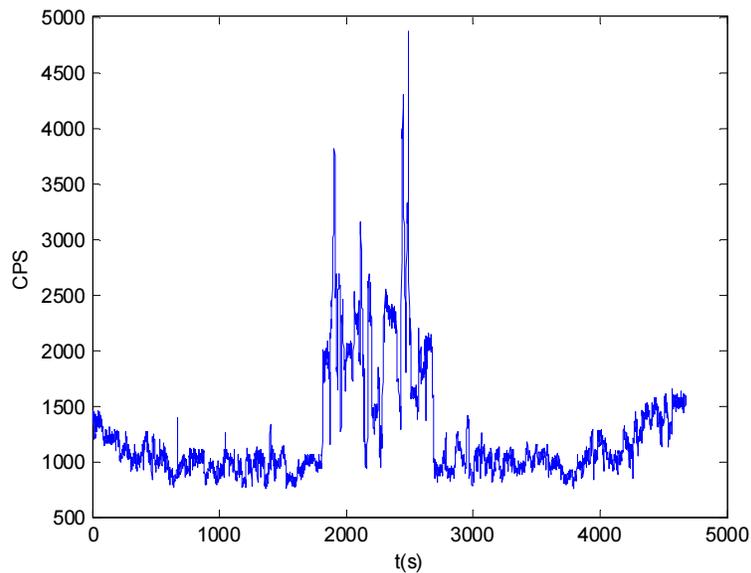

Figure 2: Histogram of Gamma Counts Rate.

### 3.2 Computational Methods Using RIID Algorithms

One set of computational tools consists of radiation isotope identification (RIID) algorithms. The operational principle of most RIID algorithms consists of matching either spectral peaks or entire measured spectra to known templates in a spectral library of isotopes, which may also include attenuated and distorted spectra of shielded sources.

Unless the source registers enough counts to display a clear spectral line, RIID algorithms fail. Spectra measured during one-second time intervals are, in general, very noisy, and do not exhibit clear spectral lines. Thus, any RIID algorithm applied to noisy and largely featureless one-second spectra would yield inconclusive results. In our prior studies, applying multiple regression method of GADRAS, Maximum Likelihood, and Fuzzy Logic RIID algorithms to one-second spectra produces a large number of false alarms and missed detections [2]. The cause for large numbers of false alarms was statistical background fluctuations, which produced random peaks in spectral channels corresponding to emission lines of isotopes of interest. As an illustration, in Figure 3 we plot four one-second spectra from the nuclear background data set measurements in downtown Chicago shown in Figures 1 and 2. The four spectra correspond to four tallest spikes in the time series in Figure 2. The spectra are noisy with poorly visible spectral lines. One can recognize K-40 line at 1440KeV, but the rest of the spectral lines are blurry.

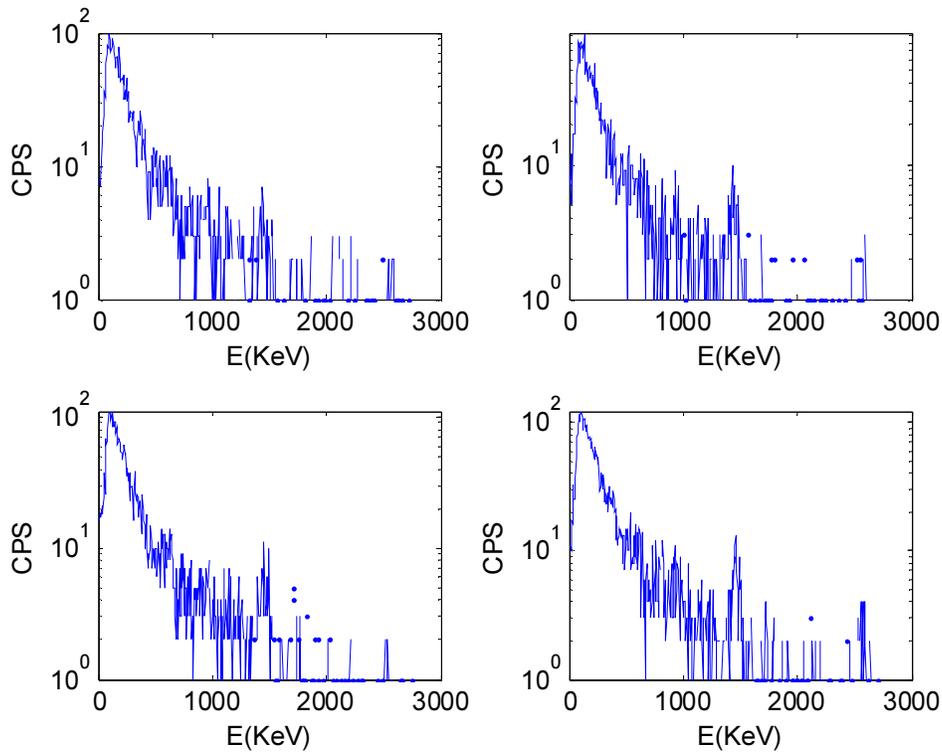

Figure 3. Four one-second spectra of the nuclear background measurements corresponding to the four tallest spikes in the time series in Figure 2.

RIID algorithm performance can be improved if spectral data S/N is increased by averaging over multiple time frames containing one-second spectra. As an illustration, in Figure 4 we plot search-time averaged number of gamma counts $<N_\gamma>_T$ as function of energy $E$ for the data in Figures 1 and 2. The resulting plot shows clear isotopic spectral lines. The fundamental problem with this approach is that the number of time frames to use in averaging is not known. Averaging the spectra over the entire search time will increase S/N of the spectral lines of the most common NORM isotopes encountered during the search. On the other hand, a source is not commonly observed repeatedly during the entire search. Thus, source spectral lines will be washed out by averaging with a large number of time frames containing random background fluctuations in the same spectral bands.

One possibility to address the spectrum averaging problem is to use a moving window average of different lengths. This approach is, generally, incorrect because there is a possibility that the same source could be measured by the detector in frames which are not sequential in time. This can happen if during search the detector platform approaches the same source from different directions, or if the source has been moving during the screening campaign. There is no guidance on the limits of the averaging window. If one is averaging over small number of randomly selected time frames, statistical fluctuations of the background may randomly produce what appears to be a resemblance of a source spectral line. On the other hand, averaging over large number of one-second frames can lead to washing out of source spectral lines. Since there is no a-priori information, one can average over randomly selected combinations of one-second spectra. Given the large size of the data, the solution of such a combinatorial problem would require enormous computational resources.

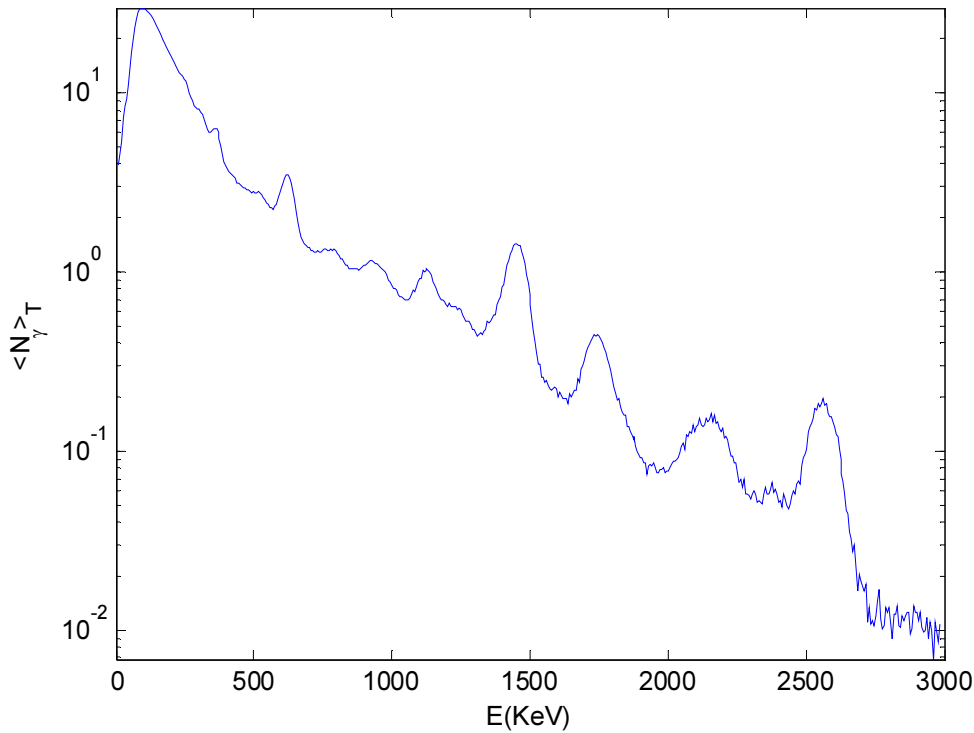

Figure 4. Search-time averaged number of gamma counts $<N_\gamma>_T$ as a function of energy $E$ for the data in Figures 1 and 2.

## 4. Data analysis using Dynamic Quantum Clustering

Our approach provides the scientific basis for selecting time frames containing one-second spectra, which can be subsequently averaged without losing information about the source. The physical idea behind this approach is that the presence of a source would not only impact a narrow spectral band by corresponding to emission spectral line, but also has the effect on the entire detector spectrum via the Compton continuum and detector response function (DRF). This allows characterizing the spectrum using multiple data features, as opposed to the single feature of spectral line. Using multiple but weakly pronounced features becomes essential to characterizing spectra with low S/N. Implementation of our method is based on the realization that one second spectra with low S/N can be grouped by overall spectral shape similarity using a data clustering technique called Dynamic Quantum Clustering (DQC). The spectra that end up in the same cluster – as determined by their overall shape - can then be averaged to enhance S/N of the isotopic spectral line. This would allow for increased accuracy of isotopic identification.

DQC uses ideas borrowed from quantum mechanics to solve the problem of clustering data – or equivalently, revealing hidden correlations among the many features being measured. The method works by creating a quantum potential that serves as a faithful proxy of the density of the data in feature space. It then uses the mathematics of quantum evolution in Hilbert space to reveal the correlated subsets of the data as simple (point-like) clusters, or as extended, parameterizable shapes. The outcome of a DQC analysis is a movie that shows how and why sets of data-points are eventually classified as members of simple clusters or as members of extended structures. Prior studies have shown that DQC has much higher sensitivity to density variation in the data compared to other clustering methods, such as the Parzen window estimator, the Support Vector machine, or K-means clustering [5-7].

Validity of the proposed approach was demonstrated with a proof-of-concept numerical experiment based on the example case listed above. In this study, DQC was applied to a database of 7675 one-second spectra collected with NaI(Tl) detector with 512 spectral channels in a drive-through campaign in the Chicago area. A subset of this data was displayed in Figures 1 and 2. Approximately 70 one-second spectra in the total dataset consisted of measurements with a Cs-137 and Co-60 radioactive sources with various levels of S/N. Cs-137 isotope has a characteristic line at 667KeV, while Co-60 has two characteristic lines at 1173KeV and 1332KeV. Search time-averaged number of gamma counts $<N_\gamma>_T$ for all 7765 spectral measurements in the data set is plotted in Figure 5. Note that the isotopic spectrum is essentially the same as that in Figure 4. There are no visible Cs-137 or Co-60 lines. This is highlighted by drawing vertical dashed lines in Figure 5 to indicate spectral locations of where the centers of Cs-137 and Co-60 lines are expected to be. In addition, there is a spectral line of NORM Bi-214 centered at 609KeV, which can potentially obscure the Cs line.

Using DQC analysis, we clustered the data in the feature space of 120 singular value decomposition (SVD) components. We have observed that two point-like clusters had a clear separation from the rest of the clusters in the feature space. These stand-out clusters represent anomalies in the data. Figures 6(a)-(c) display the results of clustering in the space of first three SVD components. The anomaly clusters are circled red dots, while the rest of the data is displayed as aqua-marine dots. Frame captures in Figures 6(a), (b) and (c) show that as the field of view is rotated, anomaly clusters 1 and 2 are moving relative to the background points. Since all data clusters to points, the anomaly clusters 1 and 2 were extracted algorithmically by putting a sphere around each point of radius 0.1 in the feature space.

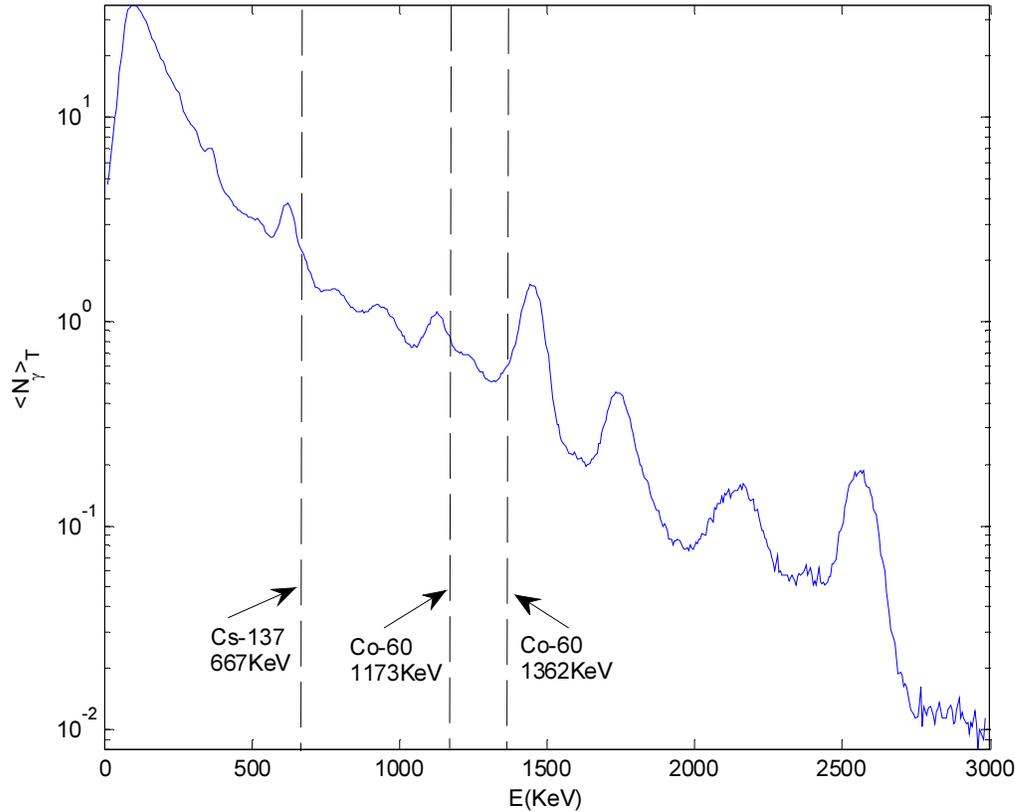

Figure 5. Time-averaged number of gamma counts $<N_\gamma>_T$ for 7675 one-second spectra consisting of background measurements and approximately 70 one-second spectra of Cs-137 and Co-60 source measurements. Vertical dashed lines indicate where the centers of Cs-137 and Co-60 spectral lines are expected to be.

Upon inspection of the two anomaly clusters, we have observed that both clusters consist of spectra containing the Cs-137 and Co-60 signals and no background-only spectra. Also, we have confirmed that all spectra in the original set which could contain Cs-137 or Co-60 or both signals have clustered into either one of the anomaly clusters. One anomaly cluster contains 24 spectra and has a strong Cs and Co peaks in every spectrum. The second anomaly cluster contains 44 spectra and shows modest Cs peaks, or arguably Cs peaks that are not visible. Some of the spectra in the second cluster also show modest Co-60 peaks. One-second spectra in clusters 1 and 2 are shown in Figures 7 and 8, respectively. Each subplot has log-linear axes, where the x-axis is the energy $E$ in units of KeV, and the y-axis is gamma counts per second. The energy range was limited to 2000KeV because the counts above that threshold value were negligible. Central locations of Cs-137 and Co-60 spectral peaks are marked by vertical dashed lines.

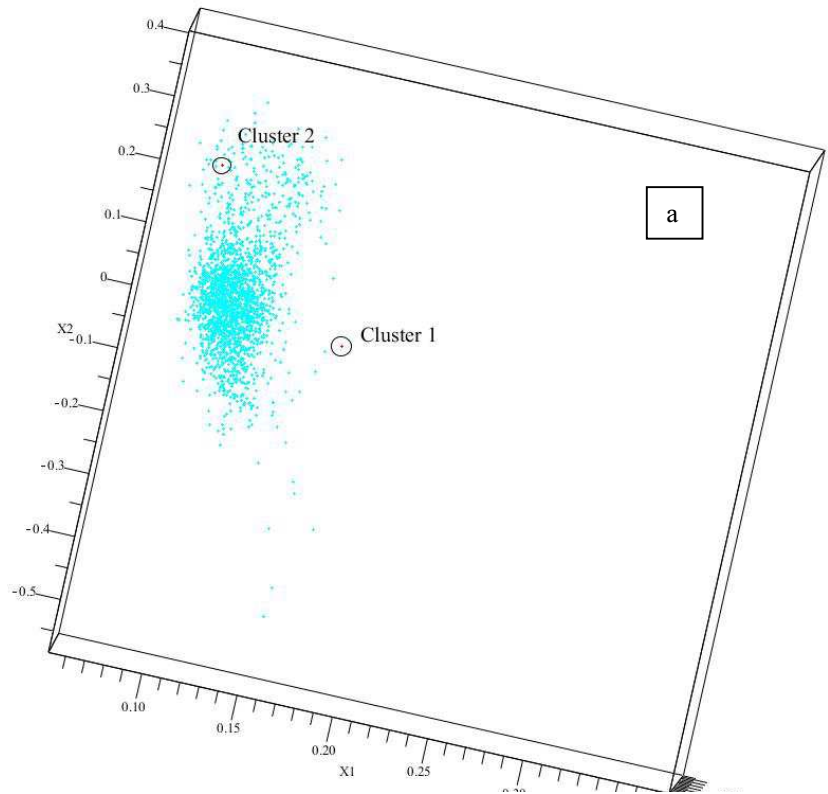
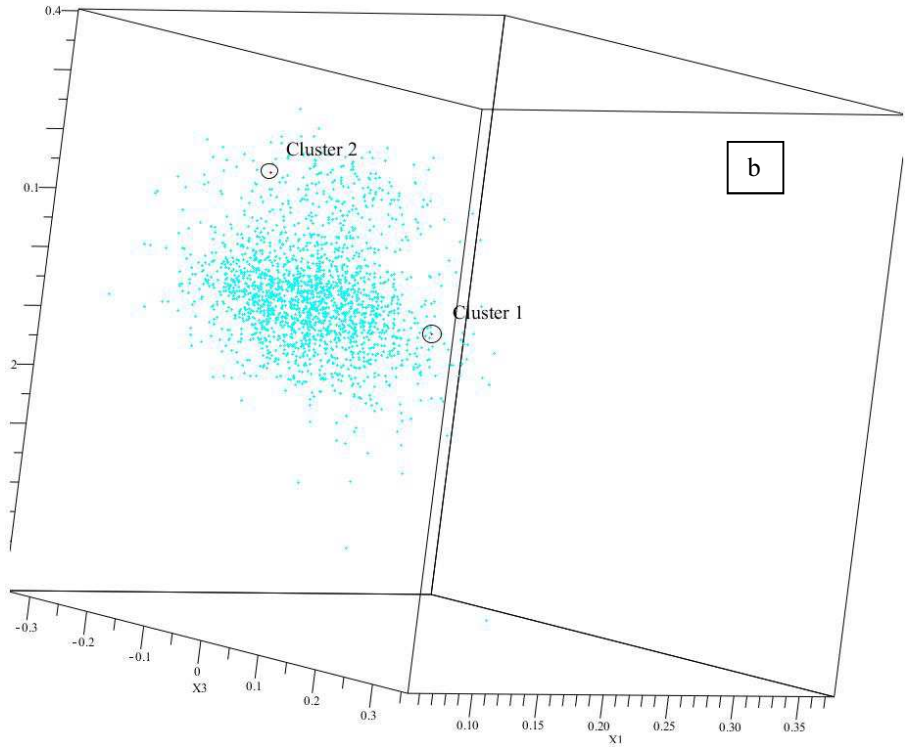

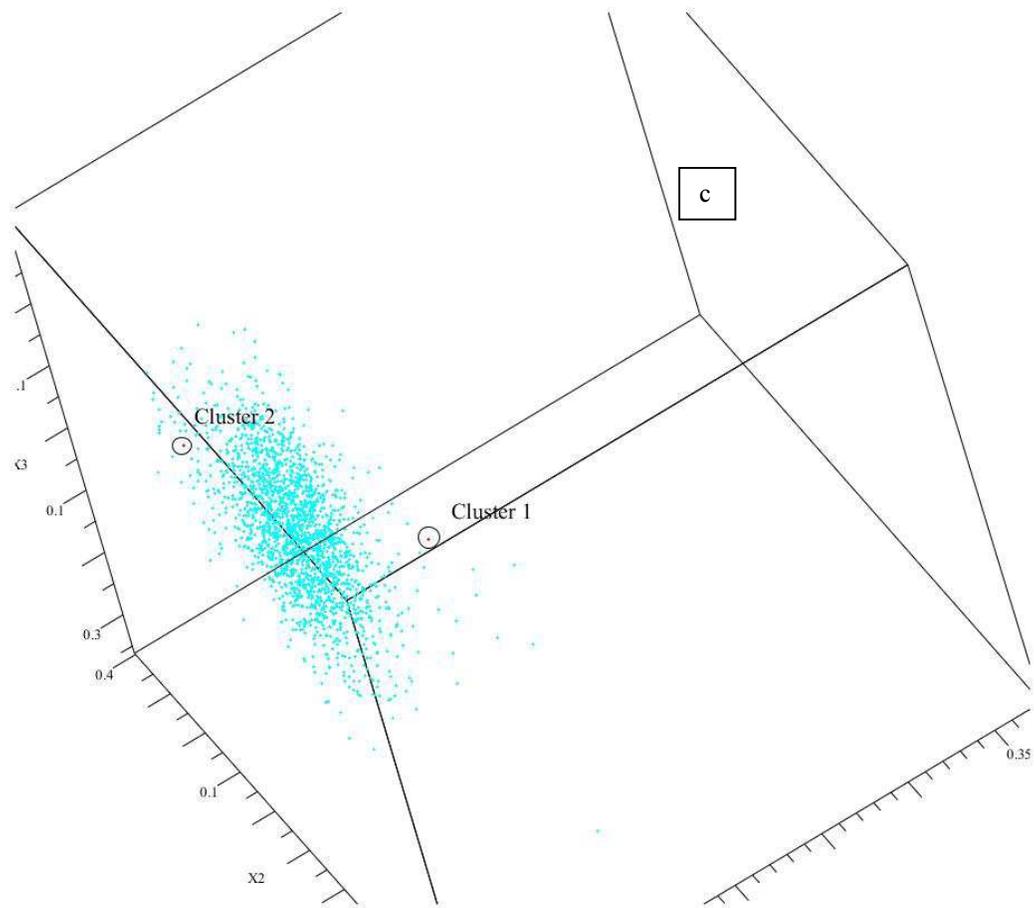

Figure 6. Separation of two clusters containing Cs-137 and Co-60 spectra (circled red dots) from background data (aqua-marine dots) in the feature space of the first three SVD components. Frame captures in (a), (b) and (c) show that as the field of view is rotated, anomaly clusters 1 and 2 are moving relative to the background points.

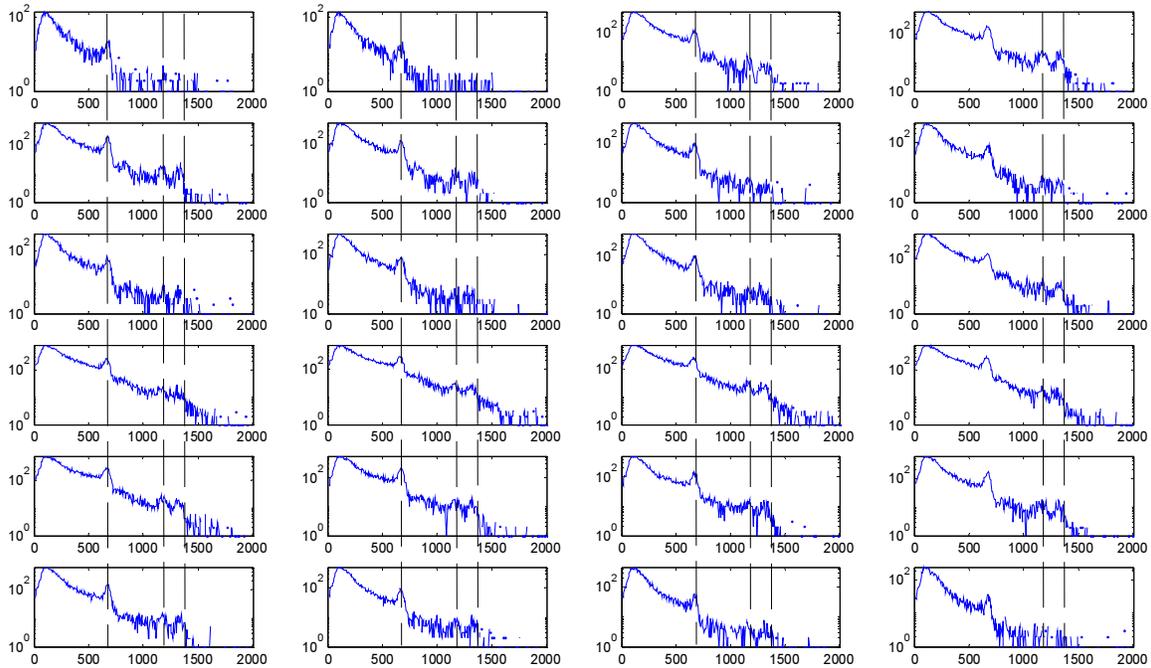

Figure 7. Log-linear plots of 24 one-second spectra displaying strong Cs peaks at 667KeV and good Co-60 spectra at 1173KeV and 1362KeV in Cluster 1. Centers of the Cs-137 and Co-60 peaks are marked by vertical dashed lines.

Seven spectra in the second cluster each have total gamma counts less than 4000. These spectra might not be registered by threshold-triggered alarm in urban search, where the total count of the background fluctuates between 1000 and 5000 total gamma counts per second (according to Figure 2). In Figure 8, the spectra with total counts less than 4000 are enclosed by red boxes. It should also be noted that some of one-second spectra in the second cluster with total counts exceeding 4000 do not display clear isotopic lines. For example, spectrum #39 in Figure 8 (9th row, 3rd column) has over 6000 total counts but no clear isotopic spectral lines.

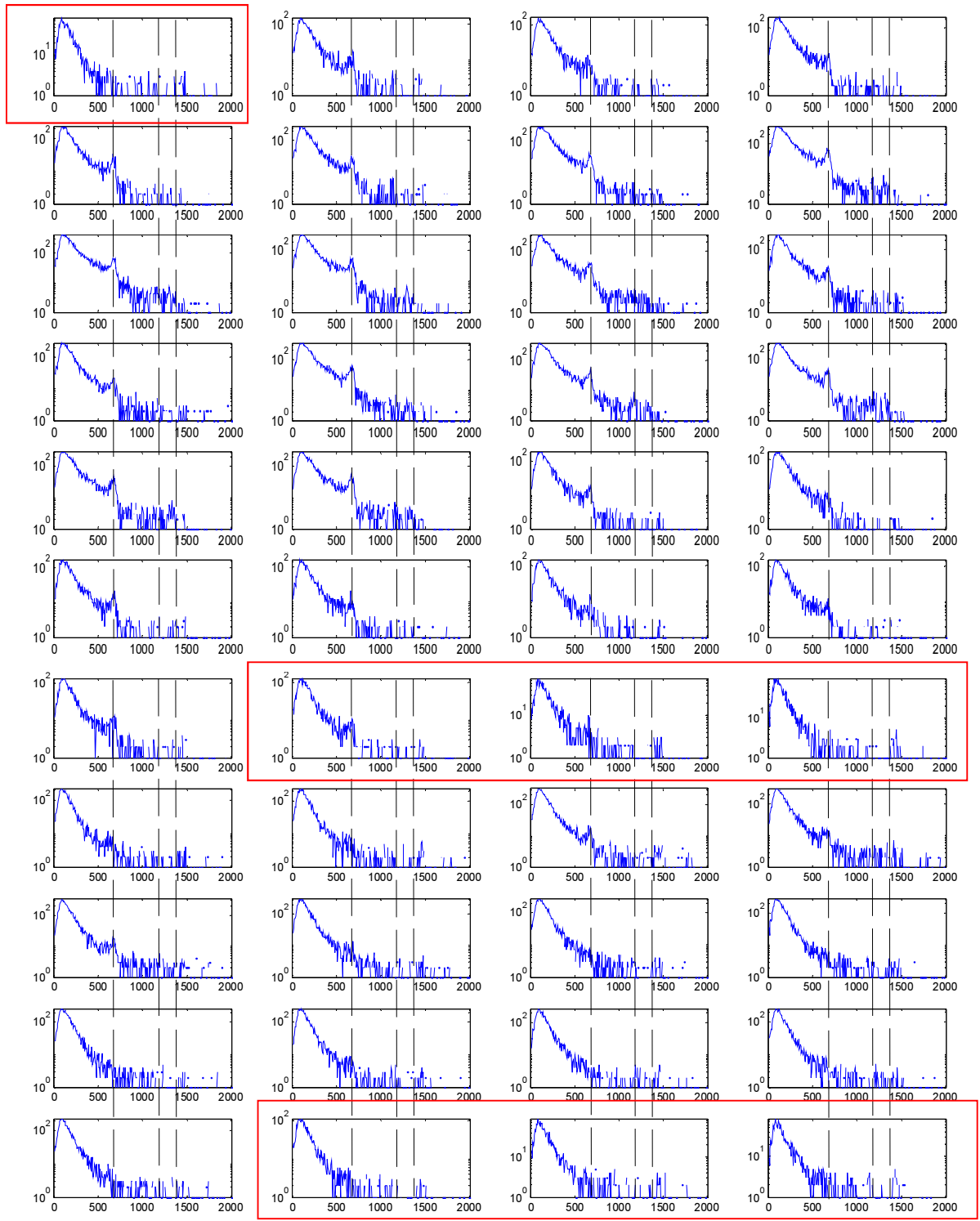

Figure 8. Log-linear plots of 44 one-second spectra containing modest to indistinct Cs-137 peaks at 667KeV and Co-60 spectra at 1173KeV and 1362KeV in Cluster 2. Centers of the isotopic peaks of Cs-137 and Co-60 are marked by the dashed vertical lines. Spectra with total counts less than 4000 are enclosed by red boxes.

As hypothesized previously, the data in anomaly clusters can be integrated in time to improve source(s) S/N without losing spectral information in the averaging process. Figure 9 shows log-linear plots of cluster-averaged spectra in the first and second clusters. The average spectrum of the first anomaly cluster in Figure 9(a) shows a very strong Cs-137 peak at 667KeV, and strong Co-60 peaks at 1173KeV and 1362KeV. The average spectrum of the second anomaly cluster in Figure 9(b) shows a good Cs-137 peak and weak Co-60 peaks.

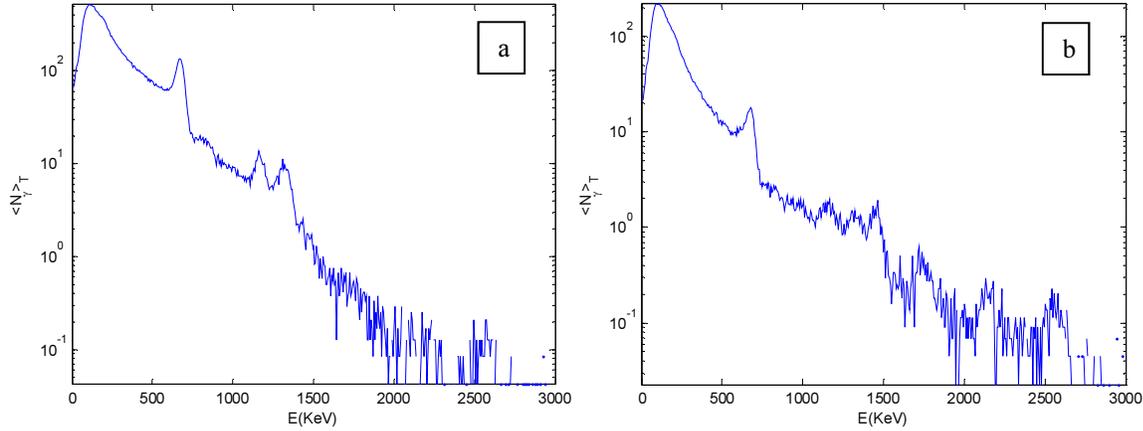

Figure 9. (a) Cluster-averaged number of gamma counts $<N_\gamma>_T$ of 24 spectra in cluster 1. (b) Cluster-averaged number of gamma counts $<N_\gamma>_T$ of 44 spectra in cluster 2. In either (a) or (b), there is a clearly identifiable Cs peak at 667KeV. The spectrum in (a) shows clear Co-60 peaks at 1173 and 1362KeV, while the spectrum in (b) shows weak Co-60 peaks

In the present study, one-second spectra of Cs-137 and Co-60 sources were measured in a time sequence. The principle of DQC operation is such that clustering is insensitive to temporal correlations of cluster members. That is, the same result as in the present study would be obtained if the one-second spectra with Cs-137 and Co-60 sources were randomly distributed among the entire database of 7675 spectra. This feature of DQC allows using the method for detecting and tracking sources in real time.

5. **Future Work**

Development of DQC-based technique for detection of sources will require further algorithm optimization studies and large-scale validation and verification experiments. We will investigate the relationship between sensitivity of DQC feature identification in the spectrum to the detector response function (DRF) of NaI(Tl)). This will provide guidelines on limits of source detection capability of DQC-based technique. It is expected that DQC performance will depend on such factors as source spectrum and signal strength, background isotopic composition and variability with time. This study will also provide an indication if the detector spectral channels can be ranked in order of importance for DQC analysis. Channels with least importance could be excluded from data to reduce its size, and hence increase the speed of analysis. In the preliminary study, DQC analysis was performed in 120-dimensional feature space. We will study the dependence of DQC performance on dimension of the feature space for different sources and backgrounds. In the present study, we considered clustering of spectral data taken at different times. In the future work, we will investigate clustering of temporal data in the energy domain, and compare the information content of either approach. We will also investigate the possibilities of expediting DQC performance. For example, it has been shown recently that DQC can be parallelized and efficiently executed using graphics processors [8]. We will investigate if faster algorithm execution can be achieved by clustering of spectral data in time or clustering of temporal data in the energy domain by introducing appropriate simplifications in either approach. DQC performance on spectral data collected with other medium resolution detectors used in search applications will be considered as well. Also, we will investigate the possibilities for developing a technique using DQC for real-time source detection and tracking combined with an estimation approach, i.e. building on the data set as it is being collected.

## 6. Conclusion

We have designed a method for detecting nuclear sources in search applications using clustering of spectral data. Our method would allow for increased accuracy of isotopic identification and lower false alarm rate. Validity of our approach has been demonstrated in a proof-of-principle study using Dynamic Quantum Clustering (DQC) method. Validation study consisted of a data set of spectra measured in one-second intervals with Sodium Iodide detector. The data set of over 7000 spectra consisted mostly of urban background measurements, and approximately 70 measurements of Cs-137 and Co-60 sources. Using DQC analysis, we have observed that all spectra containing the Cs-137 and Co-60 signal cluster away from the background into two point-like clusters. One cluster contains spectra with strong Cs and Co peaks in every spectrum. The second cluster contains spectra with modest Cs and Co peaks, or arguably Cs and Co peaks that are not visible. Averages of spectra in either cluster reveal clear Cs-137 and Co-60 spectral lines.

**Acknowledgement**

This work was supported in part by the National Science Foundation Cyber-Physical Systems Program under Work for Others Agreement # P-13025. The submitted manuscript has been created by UChicago Argonne, LLC, Operator of Argonne National Laboratory ("Argonne"). Argonne, a U.S. Department of Energy Office of Science laboratory, is operated under Contract No. DE-AC02-06CH11357.